\documentclass[conference]{IEEEtran}

\usepackage[T1]{fontenc}
\usepackage[utf8]{inputenc}
\usepackage{cite}
\usepackage{amsmath,amssymb}
\usepackage{booktabs}
\usepackage{array}
\usepackage{url}
\usepackage{hyperref}
\hypersetup{colorlinks=true, linkcolor=black, citecolor=black, urlcolor=blue}

\usepackage{pgfplots}
\pgfplotsset{compat=1.18}
\usepackage{pgfplotstable}

\usepackage{tikz}
\usetikzlibrary{shapes.geometric, arrows.meta, positioning, fit, backgrounds}

\begin{document}

\title{A Contextual Help Browser Extension to Assist\\
Digital Illiterate Internet Users}

\author{
  \IEEEauthorblockN{Christos Koutsiaris}
  \IEEEauthorblockA{
    School of Science and Computing\\
    South East Technological University\\
    Ireland\\
    \textit{github.com/unseen1980/acro-helper}
  }
}

\maketitle

\begin{abstract}
This paper describes the design, implementation, and evaluation of a
browser extension that provides contextual help to users who hover over
technological acronyms and abbreviations on web pages. The extension
combines a curated technical dictionary with OpenAI's large language
model (LLM) to deliver on-demand definitions through lightweight
tooltip overlays. A dual-layer artificial intelligence (AI) pipeline,
comprising Google Cloud's Natural Language Processing (NLP) taxonomy
API and OpenAI's ChatGPT, classifies each visited page as
technology-related before activating the tooltip logic, thereby
reducing false-positive detections. A mixed-methods study with 25
participants evaluated the tool's effect on reading comprehension and
information-retrieval time among users with low to intermediate digital
literacy. Results show that 92\% of participants reported improved
understanding of technical terms, 96\% confirmed time savings over
manual web searches, and all participants found the tooltips
non-disruptive. Dictionary-based definitions were appended in an
average of 2{,}135~ms, compared to 16{,}429~ms for AI-generated
definitions and a mean manual search time of 17{,}200~ms per acronym.
The work demonstrates a practical, real-time approach to bridging the
digital literacy gap and points toward extending contextual
help to other domains such as medicine, law, and finance.
\end{abstract}


\section{Introduction}

The rapid growth of digital technology over the past five decades has
been driven in large part by advances in semiconductor density, as
captured by Moore's Law \cite{schaller1997}. As devices have become
smaller and more capable, technology has permeated almost every aspect
of daily life, bringing with it an ever-expanding vocabulary of
abbreviations, acronyms, and technical jargon
\cite{keats2010,gablerjan2015}.

Online content, including news portals, e-commerce sites, and government
services, increasingly contains this specialist language. For users who do not
interact with technology on a deep level, encountering unfamiliar
terms such as \textit{CPU}, \textit{SSD}, or \textit{API} can block
comprehension, reduce engagement, and even cause social or financial
harm \cite{czaja2019}. This group is commonly described as
\emph{digitally illiterate}: individuals who lack the skills, knowledge,
or confidence to use digital tools effectively \cite{leahy2009}.

Existing remedies, such as formal education programs, ICDL curricula, and
UNESCO frameworks \cite{unesco2019,icdl}, require institutional
coordination and long lead times. They do not help a user who
encounters an unknown acronym \emph{right now}, mid-sentence, on a
technology news page.

This paper presents \textbf{Acro Helper}, a Chrome browser extension
that addresses this gap by injecting inline definitions at the point of
need. The extension automatically detects the page category, identifies
technical terms, and attaches hover-activated tooltip definitions
without requiring the user to open a new tab or interrupt their reading
flow. An empirical study with 25 participants validates its
effectiveness and measures its performance.

\section{Related Work}

\subsection{Digital Illiteracy}

Digital illiteracy is broadly defined as the inability to use
technology for reading, writing, communicating, and accessing
information in a digital context \cite{leahy2009}. Martin and
Grudziecki \cite{martin2006} describe a three-level model (Digital
Competence, Digital Usage, and Digital Transformation), where the
lowest level covers foundational skills, knowledge, and attitudes. This
work targets primarily Level~I users.

Key factors associated with low digital literacy include age, education
level, and socioeconomic background \cite{niehaves2014,schaffer2007}.
Even in developed countries the problem is substantial: a report by
Accenture found that one in five Irish adults under 34 self-rates their
digital skills as average or below average \cite{accenture2020}, while
the Netherlands reports 15\% of the population lacking advanced digital
skills \cite{digitalgovnl2021}. Pew Research Center projections suggest
the trend toward a more technology-driven world will deepen these
inequalities without intervention \cite{anderson2021}.

\subsection{Acronyms as Barriers to Comprehension}

Shulman et al.\ \cite{shulman2020} demonstrated that exposure to jargon
reduces readers' interest in similar articles compared with jargon-free
equivalents. Appelman \cite{appelman2019,appelman2021} showed that
readers develop negative sentiments towards acronyms they do not
understand, and that such exposure can discourage further reading of
technology-related content. For users who already have low digital
skills, this creates a feedback loop that prevents knowledge
acquisition.

\subsection{Contextual Help Tools}

Several prior works have built contextual help systems for users with
limited digital skills:
\begin{itemize}
  \item \textbf{Tipper} \cite{dai2015} is a browser extension providing
    contextual help on icons, links, and buttons; usability tests with
    seniors were strongly positive.
  \item \textbf{LemonAid} \cite{chilana2012} is a crowdsourced,
    selection-based help system for web applications, achieving a
    result for 90\% of help requests.
  \item Yeh et al.\ \cite{yeh2011} demonstrated screenshot-based
    contextual help for desktop GUIs, validated across 60 real tasks.
  \item Yadav et al.\ \cite{yadav2015} showed that SMS-based acronym
    retrieval can assist semi-literate users outside browser
    environments.
  \item \textbf{NASA Acronyms} \cite{nasa2022} is an open-source
    browser extension with over 25{,}000 space-related acronym
    definitions, available in Chrome and Firefox.
\end{itemize}

None of these prior tools combine AI-driven page classification with
LLM-powered, context-aware acronym definition delivery in a fully
automated pipeline.

\subsection{NLP for Web Content Classification}

Atkinson and Van der Goot \cite{atkinson2009} showed that text-mining
techniques applied to news articles can detect content categories in
near real time. Tokenization, stemming, and lemmatization
\cite{balakrishnan2014} improve classification accuracy during
preprocessing. Lämmel \cite{lammel2007} demonstrated that XPath-style
selectors improve DOM parsing performance. This body of work underpins
the content-extraction and classification pipeline used in Acro Helper.

\section{Research Questions}

Three research questions guide this work:

\begin{description}
  \item[RQ1] Can a contextual help browser extension improve the
    comprehension of online articles containing technical terms among
    users with low digital literacy?
  \item[RQ2] Does the contextual help tool positively affect
    information-retrieval time for these users?
  \item[RQ3] How can AI assist a browser extension in accurately
    classifying web pages by content domain?
\end{description}

The central hypothesis is that showing on-demand definitions,
activated by mouse hover, eliminates the need to leave the active tab,
increases comprehension, and reduces the total time spent resolving
unknown terms.

\section{System Architecture}

\subsection{Extension Building Blocks}

Acro Helper is built with the \textbf{Plasmo framework}
\cite{plasmo}, which enforces best-practice folder structures,
provides TypeScript support, and manages bundling. The extension
follows the standard Chrome extension model, composed of:

\begin{itemize}
  \item \textbf{Content script} — runs in the context of every visited
    page, reads and modifies the DOM.
  \item \textbf{Background service worker} — listens for browser events
    and routes messages.
  \item \textbf{Manifest v3 file} — declares permissions, entry points,
    and metadata.
  \item \textbf{Tab page} — a bundled React page providing a
    full-text dictionary search interface.
\end{itemize}

All content-manipulation logic runs asynchronously and non-blocking,
so the browser's normal rendering pipeline is not affected.

\subsection{Four-Phase Processing Pipeline}

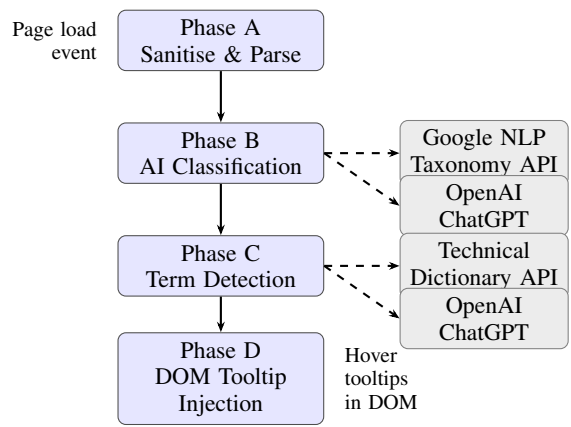
\begin{figure}[htbp]
\centering
\begin{tikzpicture}[
  font=\small,
  box/.style={rectangle, rounded corners=3pt, draw=black!70,
              fill=blue!10, minimum width=2.6cm, minimum height=0.7cm,
              text centered, text width=2.5cm},
  ext/.style={rectangle, rounded corners=3pt, draw=black!50,
              fill=gray!15, minimum width=2.1cm, minimum height=0.6cm,
              text centered, text width=2.0cm},
  arr/.style={-{Stealth[length=4pt]}, thick},
  phase/.style={font=\bfseries\small, fill=white}
]
\node[box] (A) at (0,0)   {Phase A\\Sanitise \& Parse};
\node[box] (B) at (0,-1.5){Phase B\\AI Classification};
\node[box] (C) at (0,-3.0){Phase C\\Term Detection};
\node[box] (D) at (0,-4.5){Phase D\\DOM Tooltip Injection};

\node[ext] (goog) at (3.5,-1.5) {Google NLP\\Taxonomy API};
\node[ext] (oai1) at (3.5,-2.2) {OpenAI\\ChatGPT};
\node[ext] (dict) at (3.5,-3.0) {Technical\\Dictionary API};
\node[ext] (oai2) at (3.5,-3.7) {OpenAI\\ChatGPT};

\draw[arr] (A) -- (B);
\draw[arr] (B) -- (C);
\draw[arr] (C) -- (D);

\draw[arr,dashed] (B.east) -- (goog.west);
\draw[arr,dashed] (B.east) -- (oai1.west);
\draw[arr,dashed] (C.east) -- (dict.west);
\draw[arr,dashed] (C.east) -- (oai2.west);

\node[left=0.15cm of A, text width=1.5cm, font=\footnotesize, align=right]
  {Page load\\event};
\node[right=0.15cm of D, text width=1.8cm, font=\footnotesize]
  {Hover\\tooltips\\in DOM};
\end{tikzpicture}
\caption{Four-phase pipeline of the Acro Helper browser extension.
  Dashed arrows indicate calls to external AI services.}
\label{fig:architecture}
\end{figure}

When a page finishes loading, the extension executes four sequential
phases, illustrated in Fig.~\ref{fig:architecture}.

\textbf{Phase A — Content Sanitisation and Parsing.}
The extension clones the document and strips noise elements (headers,
footers, cookie banners, navigation menus, sidebars, and
\texttt{<script>} tags) using \texttt{querySelectorAll} with targeted
CSS selectors. The Mozilla Readability library \cite{readability} then
extracts the main article body, and a whitespace-normaliser collapses
the result into clean, continuous text.

\textbf{Phase B — AI-Assisted Page Classification.}
The cleaned text is sent to the Google Cloud Natural Language API's
Content Classification endpoint. If the returned categories include any
technology-related entry (e.g., ``Computers'', ``Software'',
``Internet'', ``Engineering \& Technology'') the page is classified as
tech-related. If not, the content is forwarded to OpenAI's ChatGPT with
a boolean prompt: \emph{``Does this text contain technology-related
terms?''}. This dual-layer approach handles edge cases where
technology terms appear within a non-technology context (e.g.,
blockchain policy articles classified as ``Politics'').

\textbf{Phase C — Term Detection.}
If the page is classified as technology-related, the extension queries
its remote dictionary server for a list of known acronyms (keys only,
without definitions, to minimise payload). A case-insensitive
word-boundary regular expression (\texttt{/\textbackslash b\{key\}\textbackslash b/gi}) is applied
to the article text. For each matched term, the definition is fetched
on demand. A secondary OpenAI pass identifies any additional acronyms
that are not present in the dictionary, returning contextually accurate
definitions.

\textbf{Phase D — DOM Tooltip Injection.}
Detected acronyms in the live DOM are wrapped in
\texttt{<dfn><abbr title="...">} tags. The \texttt{title} attribute
carries the definition; browsers natively render it as a tooltip on
hover. A custom replacer function targets only text nodes, avoiding
interference with HTML attributes or existing hyperlinks.

\subsection{Dictionary Search Tab Page}

A dedicated tab page (accessible from the extension popup) allows
users to search the full acronym dictionary in real time. Data is
fetched asynchronously and filtered client-side on each keystroke,
giving instant results. This feature is useful when users encounter an
unfamiliar term outside the browser (e.g., printed material or
in-person conversation) and wish to look it up.

\section{Methodology}

\subsection{Research Design}

A mixed-methods \emph{Explanatory Sequential} design was adopted
\cite{toyon2021}: quantitative data collection (interactive survey and
performance benchmarks) was followed by qualitative data collection
(interviews), with the qualitative phase used to explain quantitative
findings.

\subsection{Participants}

A total of 25 participants were recruited across three groups over four
weeks (6~April -- 4~May~2023):
\begin{enumerate}
  \item \textbf{River Valley Community Centre}, North Dublin — 10
    members, predominantly aged 50+.
  \item \textbf{Coláiste Íde College of Further Education}, North
    Dublin — 10 mature students.
  \item \textbf{SAP Ireland restaurant staff}, West Dublin — 5
    participants whose work does not involve daily computer use.
\end{enumerate}

Demographic summary: 60\% aged 31--59, 36\% aged 18--30, 4\% aged
60+; 52\% female, 44\% male; 52\% with only a high-school diploma.
The majority (72\%) self-rated their technology expertise as
\emph{intermediate}.

\subsection{Procedure}

Each session followed five steps:
\begin{enumerate}
  \item Briefing on the study and the problem being addressed.
  \item Reading a technology news article (MacBook Pro review) \emph{without} the extension enabled.
  \item Completing a pre-use questionnaire: demographics, technology
    usage, digital literacy self-assessment, and baseline experience
    with technical acronyms.
  \item Enabling the extension, re-reading the article, and interacting
    with the tooltips.
  \item Completing a post-use questionnaire assessing comprehension
    improvement, UX, and time savings; then searching for the
    definition of ``CPU'' on Google with a stopwatch to record manual
    search time.
\end{enumerate}

Post-session, each participant took part in an individual interview
analysed via thematic analysis.

\subsection{Performance Benchmarking}

The browser's \texttt{performance.now()} API was used to measure
execution time from the page-load event to the completion of tooltip
injection, across 10 independent runs for both the dictionary path and
the OpenAI path. IBM SPSS was used for descriptive statistics.

\section{Results}

\subsection{Comprehension Improvement}

\begin{figure}[htbp]
\centering
\begin{tikzpicture}
\begin{axis}[
  ybar,
  bar width=18pt,
  width=0.92\columnwidth,
  height=5cm,
  symbolic x coords={
    {Significantly},
    {Moderately},
    {Slightly},
    {No change},
    {Worse}},
  xtick=data,
  x tick label style={font=\footnotesize, rotate=20, anchor=east},
  ymin=0, ymax=60,
  ylabel={\% of participants},
  ylabel style={font=\footnotesize},
  ymajorgrids=true,
  grid style=dashed,
  nodes near coords,
  nodes near coords style={font=\footnotesize},
  enlarge x limits=0.18,
  title style={font=\small},
  title={Extension Impact on Comprehension (Q8, $n{=}25$)},
]
\addplot[fill=blue!55] coordinates {
  ({Significantly},48)
  ({Moderately},44)
  ({Slightly},8)
  ({No change},0)
  ({Worse},0)
};
\end{axis}
\end{tikzpicture}
\caption{Self-reported improvement in understanding technical terms
  after using the extension. 92\% reported at least moderate improvement.}
\label{fig:comprehension}
\end{figure}
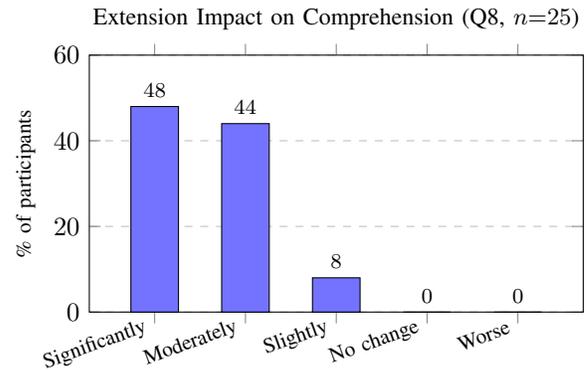

Question~8 asked: \textit{``To what extent did the extension improve
your understanding of technical terms and acronyms?''} Results are
shown in Fig.~\ref{fig:comprehension}. No participant reported a
negative or neutral outcome; 92\% reported at least moderate improvement.

\subsection{Time Savings}

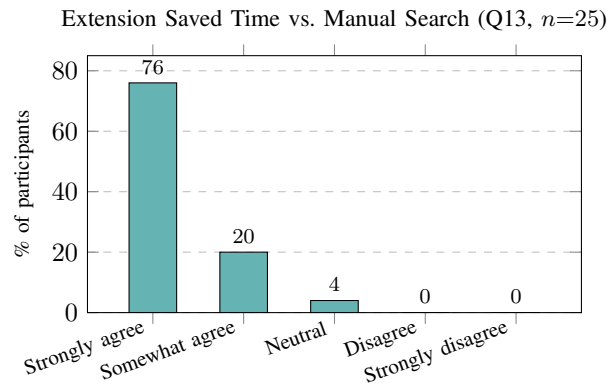
\begin{figure}[htbp]
\centering
\begin{tikzpicture}
\begin{axis}[
  ybar,
  bar width=18pt,
  width=0.92\columnwidth,
  height=5cm,
  symbolic x coords={
    {Strongly agree},
    {Somewhat agree},
    {Neutral},
    {Disagree},
    {Strongly disagree}},
  xtick=data,
  x tick label style={font=\footnotesize, rotate=20, anchor=east},
  ymin=0, ymax=85,
  ylabel={\% of participants},
  ylabel style={font=\footnotesize},
  ymajorgrids=true,
  grid style=dashed,
  nodes near coords,
  nodes near coords style={font=\footnotesize},
  enlarge x limits=0.18,
  title style={font=\small},
  title={Extension Saved Time vs.\ Manual Search (Q13, $n{=}25$)},
]
\addplot[fill=teal!60] coordinates {
  ({Strongly agree},76)
  ({Somewhat agree},20)
  ({Neutral},4)
  ({Disagree},0)
  ({Strongly disagree},0)
};
\end{axis}
\end{tikzpicture}
\caption{96\% of participants confirmed the extension saved time
  compared with searching a search engine manually.}
\label{fig:timesaving}
\end{figure}

Question~13 asked whether the extension saved time compared to manually
searching for definitions. Results are shown in Fig.~\ref{fig:timesaving};
96\% of participants agreed or strongly agreed.

\subsection{Recommendation Likelihood}

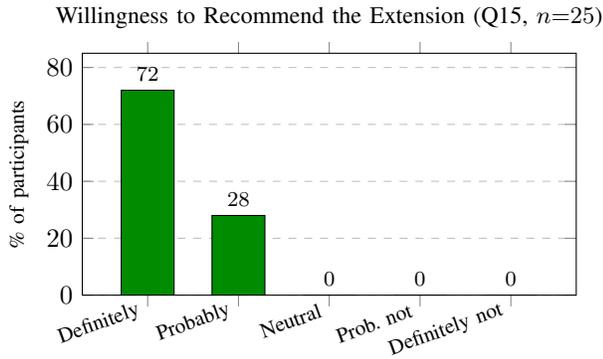
\begin{figure}[htbp]
\centering
\begin{tikzpicture}
\begin{axis}[
  ybar,
  bar width=20pt,
  width=0.92\columnwidth,
  height=4.8cm,
  symbolic x coords={
    {Definitely},
    {Probably},
    {Neutral},
    {Prob.\ not},
    {Definitely not}},
  xtick=data,
  x tick label style={font=\footnotesize, rotate=20, anchor=east},
  ymin=0, ymax=85,
  ylabel={\% of participants},
  ylabel style={font=\footnotesize},
  ymajorgrids=true,
  grid style=dashed,
  nodes near coords,
  nodes near coords style={font=\footnotesize},
  enlarge x limits=0.18,
  title style={font=\small},
  title={Willingness to Recommend the Extension (Q15, $n{=}25$)},
]
\addplot[fill=green!55!black] coordinates {
  ({Definitely},72)
  ({Probably},28)
  ({Neutral},0)
  ({Prob.\ not},0)
  ({Definitely not},0)
};
\end{axis}
\end{tikzpicture}
\caption{All participants expressed willingness to recommend the
  extension; none were neutral or negative.}
\label{fig:recommend}
\end{figure}

All 25 participants said they would recommend the extension to others
(Fig.~\ref{fig:recommend}): 72\% \emph{definitely} and 28\%
\emph{probably}.

\subsection{Manual Search Time}

Each participant searched for the definition of ``CPU'' using Google
while the researcher used a stopwatch. Table~\ref{tab:manualsearch}
summarises the results. The mean search time was 17.2~seconds
(17{,}200~ms), with a range of 22~seconds.

\begin{table}[htbp]
\caption{Manual Search Time for the ``CPU'' Acronym ($n{=}25$)}
\label{tab:manualsearch}
\centering
\begin{tabular}{lr}
\toprule
\textbf{Statistic} & \textbf{Value (seconds)} \\
\midrule
$N$ (valid) & 25 \\
Mean        & 17.20 \\
Median      & 16.00 \\
Std.\ deviation & 4.72 \\
Minimum     & 12.00 \\
Maximum     & 34.00 \\
Range       & 22.00 \\
\bottomrule
\end{tabular}
\end{table}

\subsection{Extension Performance}

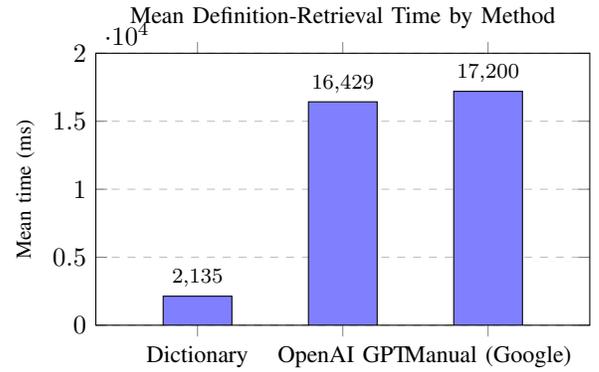
\begin{figure}[htbp]
\centering
\begin{tikzpicture}
\begin{axis}[
  ybar,
  bar width=26pt,
  width=0.92\columnwidth,
  height=5.2cm,
  symbolic x coords={{Dictionary},{OpenAI GPT},{Manual (Google)}},
  xtick=data,
  x tick label style={font=\small},
  ymin=0, ymax=20000,
  ylabel={Mean time (ms)},
  ylabel style={font=\footnotesize},
  ymajorgrids=true,
  grid style=dashed,
  nodes near coords,
  nodes near coords style={font=\footnotesize},
  enlarge x limits=0.35,
  title style={font=\small},
  title={Mean Definition-Retrieval Time by Method},
]
\addplot[fill=blue!50] coordinates {
  ({Dictionary},2135)
  ({OpenAI GPT},16429)
  ({Manual (Google)},17200)
};
\end{axis}
\end{tikzpicture}
\caption{Mean time (ms) to obtain a definition via the dictionary,
  via OpenAI ChatGPT, and via a manual Google search. Lower is better.
  Both automated methods outperform or match manual search.}
\label{fig:performance}
\end{figure}

Table~\ref{tab:performance} and Fig.~\ref{fig:performance} compare the
three definition-delivery methods across 10 benchmark runs. The
dictionary path ($\mu = 2{,}135$~ms) is approximately 7.7$\times$
faster than the AI path ($\mu = 16{,}429$~ms), and both are
substantially faster than the mean manual search time of
17{,}200~ms.

\begin{table}[htbp]
\caption{Extension Execution Time vs.\ Manual Search (ms, $n{=}10$ runs)}
\label{tab:performance}
\centering
\begin{tabular}{lrrrrr}
\toprule
\textbf{Method} & \textbf{Mean} & \textbf{Median} & \textbf{SD} & \textbf{Min} & \textbf{Max} \\
\midrule
Dictionary & 2{,}135 & 2{,}116 & 178 & 1{,}907 & 2{,}421 \\
OpenAI GPT & 16{,}429 & 16{,}431 & 3{,}578 & 10{,}568 & 22{,}473 \\
Manual (Google) & 17{,}200 & 16{,}000 & 4{,}717 & 12{,}000 & 34{,}000 \\
\bottomrule
\end{tabular}
\end{table}

\subsection{User Experience}

\begin{itemize}
  \item \textbf{Tooltip distraction (Q9):} 100\% of participants
    answered \emph{No}: the tooltips did not distract from reading.
  \item \textbf{Definition clarity (Q10):} 88\% agreed the definitions
    were clear (44\% strongly, 44\% somewhat); 12\% were neutral.
  \item \textbf{Website performance impact (Q11):} 96\% noticed no
    slowdown; only 1 participant reported a perceived impact.
  \item \textbf{Inclination to explore more content (Q12):} 72\%
    agreed the tool made them more likely to read additional
    technology articles.
\end{itemize}

\subsection{Qualitative Findings}

Thematic analysis of 25 individual interviews produced three
recurrent themes:

\textbf{Theme 1 — Clarity and Depth of Definitions.}
Approximately 70\% of participants preferred AI-generated definitions
because they were more concise and contextually relevant; dictionary
definitions were seen as more detailed but sometimes verbose.
Participants requested a tiered definition feature (brief vs.\
extended).

\textbf{Theme 2 — User Interface.}
The tooltip trigger mechanism (underline + question-mark cursor) was
widely understood. Users requested customisable font size and colour.
One participant suggested a two-part tooltip: the expanded acronym as a
title, followed by the full definition.

\textbf{Theme 3 — Additional Resources.}
Most participants felt the definition alone was sufficient. A minority
requested links for further reading or, where possible, illustrative
images.

One notable qualitative case involved a participant in
his twenties studying cybersecurity who self-identified as being on the
autism spectrum and experiencing short-term memory difficulties. He
described his current workaround of writing down or screenshotting
terms before opening a new tab, a process often disrupted by forgetting the
term during the tab switch. He expressed immediate enthusiasm for Acro
Helper, noting it would ``eliminate the need to note down and manually
search for each unfamiliar term'' encountered in his studies.

\section{Discussion}

\subsection{Answers to Research Questions}

\textbf{RQ1 — Comprehension.}
The data confirm that the extension can improve comprehension of
technical online articles for users with low to intermediate digital
literacy. An overwhelming 92\% reported at least moderate improvement,
and 100\% agreed (48\% strongly) that the extension would enhance their
understanding of technical content. This is consistent with prior work showing
that inline contextual help reduces cognitive load during reading
\cite{chilana2012,dai2015}.

\textbf{RQ2 — Reading Efficiency.}
The extension improved reading \emph{efficiency} rather than raw
reading \emph{speed}: users no longer needed to context-switch to a
search engine. The dictionary path's mean of 2{,}135~ms and the
AI path's mean of 16{,}429~ms are both well below the 17{,}200~ms
mean for a single manual search; the extension handles
\emph{all} detected acronyms simultaneously, whereas each manual
search covers only one. The slight increase in total session time
observed informally reflects users' curiosity in hovering over terms
they already knew.

\textbf{RQ3 — AI-Assisted Classification.}
The dual-layer approach (Google NLP taxonomy + OpenAI fallback)
successfully addressed misclassification cases where technical
terminology appeared in non-technology contexts (e.g., blockchain
legislation). False positives from the dictionary path (for example, ``POST'' being
matched as ``Power On Self Test'' in a non-technical sentence) remain
an open problem for purely pattern-based detection
and could be resolved by embedding an on-device contextual NLP model
in a future version.

\subsection{Broader Implications}

The study highlights the potential of \emph{just-in-time} digital
literacy tools as a complement to long-term educational programs. With
96\% of participants confirming time savings and no participant
reporting a negative reading experience, such extensions can lower the
barrier to engaging with technology content without requiring prior
training. Replacing the technology taxonomy categories and the underlying dictionary
allows the same pipeline to serve medical, legal, or financial users.

\subsection{Limitations}

The sample size ($n = 25$) limits statistical power and
generalisability. Self-selection bias likely skewed the sample toward
participants already motivated to improve their digital skills. The
study was conducted in a supervised setting, which may have inflated
engagement scores relative to natural, unsupervised use.

\section{Future Work}

Based on participant feedback and benchmark results, the following
improvements are planned:
\begin{itemize}
  \item \textbf{Definition caching}: Store AI-generated definitions in
    the local dictionary after first use; subsequent visits retrieve
    from the dictionary path (2{,}135~ms vs.\ 16{,}429~ms).
  \item \textbf{Contextual NLP on-device}: Integrate a lightweight
    model to resolve polysemous acronyms (e.g., ``POST'') without a
    remote call.
  \item \textbf{Customisable UI}: Font size, colour, and tooltip layout
    controls.
  \item \textbf{Multi-language support}: Browser locale detection to
    serve definitions in the user's language via translation APIs.
  \item \textbf{Domain expansion}: Finance, medicine, and law
    dictionaries with corresponding taxonomy rules.
  \item \textbf{Contribution mechanism}: Allow users to suggest or
    correct definitions, fostering an open-source knowledge base.
\end{itemize}

\section{Conclusion}

This paper presented Acro Helper, a Chrome browser extension that uses
a hybrid dictionary and LLM pipeline to deliver real-time contextual
definitions for technical acronyms to users with low digital literacy.
A user study with 25 participants across community, education, and
corporate settings demonstrated that the tool significantly improved
comprehension (92\% of participants), saved time over manual search
(96\%), and was non-intrusive to the reading experience (100\%). The
dictionary-based path performed at 2{,}135~ms, approximately 8$\times$
faster than a single manual Google search, while AI-generated
definitions remained competitive at 16{,}429~ms and were preferred by
70\% of participants for their contextual accuracy.

The work shows that real-time, AI-assisted contextual help embedded at
the browser level is a viable and user-accepted approach to bridging the
digital literacy gap. The open-source codebase is available at
\url{https://github.com/unseen1980/acro-helper}, and a video
demonstration at \url{https://youtu.be/FiYl-itulUQ}.

\section*{Acknowledgements}
The author thanks the participants at River Valley Community Centre,
Coláiste Íde, and SAP Ireland for their time and feedback, and
Dr.\ Brenda Mullally for supervision during the original dissertation.

\bibliographystyle{IEEEtran}

\end{document}